\documentclass[journal=nalefd,manuscript=article,layout=traditional]{achemso}
\usepackage[version=3]{mhchem} % Formula subscripts using \ce{}

\author{YongJoo Kim}
\affiliation{KAIST Institute for NanoCentury, Korea Advanced Institute of Science and Technology, Daejeon, Korea, Republic of}
\author{Alfredo Alexander-Katz}
\affiliation{Department of Materials Science and Engineering, Massachusetts Institute of Technology, Cambridge, Massachusetts 02139, USA}
\email{aalexand@mit.edu}

\title
  {Anisotropic nanoparticle distribution in block copolymer model defects}

\begin{document}

\begin{abstract}

In this article, we study the thermodynamic behavior of anisotropic shape (rod and disk) nanoparticle within the block copolymer matrix by using self-consistent field theory (SCFT) simulation. In particular, we introduce various defect structures of block copolymers to precisely control the location of anisotropic particles. Different from the previous studies using spherical nanoparticles within the block copolymer model defects, anisotropic particles are aligned with preferred orientation near the defect center due to the combined effects of stretching and interfacial energy of block copolymers. Our results are important for precise controlling of anisotropic nanoparticle arrays for designing various functional nano materials.  

\end{abstract}

\section{Introduction}

In recent years, ordered array of nanoparticles (NPs) by controlled assembly has received great attention due to the inter-particle plasmonic and electric coupling between NPs \cite{art:alivisatos,art:mirkin,art:yin,art:nykypanchuk,art:devries,art:choi}. Long-range ordering of NPs can open the possibility of new functional materials for numerous applications such as SERS (Surface Enhanced Raman Scattering) spectroscopy \cite{art:xu,art:chen,art:mulvihill},  memory devices \cite{art:mukherjee,art:park,art:cui}, and solar cells \cite{art:pattantyus-abraham,art:ip,art:jean}. In addition to the 0-dimensional spherical NPs, 1-dimensional nanorods show tunable plasmonic properties depending on their geometrical arrangement \cite{art:nie,art:ng}. Furthermore, 2-dimensional nanodisks such as graphene \cite{art:ritter,art:zhuo,art:jlee} and molybdenum disulfide \cite{art:stengl,art:gopalakrishnan} quantum dots have been investigated recently due to their tunable band gap for numerous possible electronic applications. 

For the next generation NP-based device, precise localization of NP over multiple length scales is the most critical issue. Block copolymers (BCPs) have been regarded as attractive materials for the scaffold for NP array due to their ability to self-assemble into variety of structures at nanometer length scale \cite{art:leibler,art:bates,art:cpark}. By Blending NPs and BCPs, well ordered NPs are found within the self-assembled BCP matrix. It is well known that the ligand chemistry \cite{art:chiu,art:bkim} and size \cite{art:thompson,art:bockstaller1,art:kim}  of NP determine the location of NPs within the BCP matrix due to their enthalpic and entropic contributions to the blends respectively. In addition to the ligand chemistry and size of NP, conformation entropy (stretching energy) of BCPs is another important factor for the localization of NPs within the BCP matrix. For example, higher spring constant by using stiffer comb block of coil-comb BCPs helps the alignment of spherical NPs into 3d structures \cite{art:zhao,art:kao,art:kao2}, as well as the arrangement of nanorods \cite{art:thorkelsson}. Defect structures of BCP are another great example of the role of conformation entropy of BCP to localization of NPs. NPs are aggregated at curved grain boundary \cite{art:listak}, curved pattern by chemoepitaxy \cite{art:kang}, and dislocation between monolayer and double-layer of thin film \cite{art:jkim} of BCPs to minimize the conformation entropy loss. In this manner, localization of NPs at specific location would be possible, if one can design artificial defect structure of BCP at desired locations. 

Templated assisted self-assembly of BCPs is a promising method to generate controlled "defect" structure of BCPs. It is well known that control the lattice constant and shape of hydrogen silsesquioxane (HSQ) nano post array by e-beam lithography helps the formation of desired structure of highly stretched BCPs \cite{art:bita,art:yang,art:tavakkoli1,art:tavakkoli2}. Self-consistent field theory (SCFT) simulation study of templated assisted self-assembly of BCPs shows the similar result described by experiments \cite{art:mickiewicz}. Recently, blends of NPs and BCPs with various defect structures by templated assisted self-assembly are studied by using SCFT simulation \cite{art:ykim}. Free energy plots of the blends system show that the spheric NP is trapped at the defect center for every defect shape. However, when the shape of the particle deviates from the perfect sphere, we can expect that there will be favorable orientation of those particles for the different shape of BCP defect. Therefore, theoretical investigation of anisotropic shape of NP within the model defects of BCPs will be interesting problem for the both fundamental and engineering aspects.

In this article, we study blends of anisotropic (rod and disk) NPs and BCPs using SCFT simulation method with "model defect" structures of BCPs. In particular, we investigate the free energy of blends as a function of configuration of anisotropic NP near the centers of X, T, and Y shape defects of BCPs. We expect that our result will open a possibility of new way to precise control of anisotropic NP at the desired location to design new functional materials.

\section{Computation method}
\subsection{Self-Consistent Field Theory}
In this work, we simulate block copolymers (BCPs) and anisotropic (rod and disk) nanoparticles (NPs) using Self-Consistent Field Theory (SCFT) simulation. Cavity function \cite{art:sides} is introduced to describe the external particle (NPs) by excluding polymer density where the cavity function is defined. In this work, we expand this method to three dimensional system for better description of the blends of BCPs and NPs.

The effective hamiltonian used in this work including the cavity function $ \rho_{ext} $ is given by,
$$
H[ \Omega_{+}, \Omega_{-}, \{ \Omega_{ext} \}] = -CV \ln Q[ \Omega_{+}, \Omega_{-}, \{ \Omega_{ext} \} ] - iC \int d \textbf{r} \Omega_{+}\bigg( 1 - \frac{\rho_{ext}}{\rho_{0}} \bigg)  + \frac{C}{\chi N} \int d \textbf{r} \Omega_{-}^{2}
$$
where $C = \rho_{0}R_{g}^{3}  / N $ corresponds to the dimensionless concentration, $ \rho_{0} $ is the monomer concentration, $N$ is the degree of polymerization, and $R_{g}$ is radius of gyration of an ideal copolymer. $\Omega_{+}$ is interpreted as a fluctuating pressure field which couples with summation of densities of polymer A and B ($\rho_{+} = \rho_{A} + \rho_{B}$) to enforce the incompressibility condition for the local polymer density ($\rho_{+} + \rho_{ext} = \rho_{A} + \rho_{B} + \rho_{ext} = \rho_{0} $). On the other hand, $\Omega_{-} $ is interpreted as an exchange potential field that couples with local density differences between the two blocks ($ \rho_{-} = \rho_{A} - \rho_{B} $). The set of field $\{\Omega_{ext}\}$ corresponds to the set of the external field imposed on the system by the rod or disk NP which contains affinity of NPs to both blocks. The cavity function $\rho_{ext}$ describes the density of the rod or disk NP within the polymer matrix and will be discussed at next section in detail.

$Q[ \Omega_{+}, \Omega_{-}, \{ \Omega_{ext} \} ]$ is the single polymer partition function, and can be calculated as 
$$Q[ \Omega ] = \frac{1}{V}\int  d \textbf{r} q(\textbf{r},1,\Omega)   $$ 
where $ q(\textbf{r},s, \Omega) $, called the propagator, satisfies a diffusion-like equation given by,

$$ \frac{ \partial} {\partial s } q(\textbf{r},s,\Omega) = \nabla^{2}  q(\textbf{r},s,\Omega)  - \Omega(\textbf{r},s)q(\textbf{r},s,\Omega)   $$ 
having initial condition $ q(\textbf{r},0,\Omega) = 1 $. The time variable in the diffusion-like equation $s$ corresponds to the position of a polymer segment ($s = 0$ and $s = 1$ correspond to both ends of a single polymer). Therefore, the set of fields $\Omega(\textbf{r},s) $ is a function of type of polymer (A and B) described as,

$$ \Omega(\textbf{r},s)  = \bigg\{ \begin{array}{cc} i\Omega_{+}(\textbf{r}) - \Omega_{-}(\textbf{r})  - \gamma_{A}         \rho_{ext}(\textbf{r}) / \rho_{0},  \text{\ \ \ \ \ \ \ } 0 < s \leq f  \\
i\Omega_{+}(\textbf{r}) + \Omega_{-}(\textbf{r})  - \gamma_{B}\rho_{ext}(\textbf{r}) / \rho_{0}, \text{\ \ \ \ \ \ \ } f < s \leq 1
\end{array}
$$
where $f$ is the volume fraction of the A block in the block copolymer. $ \gamma_{A} $ and $ \gamma_{B} $ are the effective affinities of the nanoparticle for each block A and B. A positive value of the effective affinity yields an attractive force with the corresponding block of the BCPs. The local polymer densities $\phi_{A}$ and $\phi_{B}$ are calculated by integrating the proper propagators obtained from the diffusion-like equation. 

$$\phi_{A}(\textbf{r},\Omega)  = \frac{1}{Q}\int_{0}^{f} ds\ q^{\dagger}(\textbf{r},1-s,\Omega) q(\textbf{r},s,\Omega)$$
$$\phi_{B}(\textbf{r},\Omega)  = \frac{1}{Q}\int_{f}^{1} ds\ q^{\dagger}(\textbf{r},1-s,\Omega) q(\textbf{r},s,\Omega)$$ 
We introduce the function  $q^{\dagger}(\textbf{r},s,\Omega)$ which is equivalent to the function  $q(\textbf{r},s,\Omega)$ but the propagation along the chain starts from the B ends of the polymer. We use a Lattice Boltzman method recently developed to solve the diffusion-like equation for the propagator $q(\textbf{r},s,\Omega)$, and the program has been optimized for GPU parallel computation \cite{art:hchen}. 

To calculate the full partition function $Z = \int D\Omega e^{H[\Omega]} $, we have to integrate the over all possible configuration of the field $\Omega$. However, the SCFT of BCPs is based on the approximation of full partition function as a single partition function at the mean field (saddle point) configuration of the field $\Omega*$ described by $Z \approx e^{H[\Omega^*]}$. To find the mean field solution for $\Omega_{+}^{*}$ and $\Omega_{-}^{*}$ that satisfy the minimization of the effective Hamiltonian condition $\partial H / \partial \Omega_{+,-} |_{\Omega_{+,-}^{*}}  = 0$, a Langevin dynamic scheme is used to update $\Omega_{+}^{*}$ and $\Omega_{-}^{*}$. The computational procedure to find the mean field solution is executed as described below. First, we update the $\Omega_{-}$ field with a small Gaussian real noise to escape from metastable states, then calculate local polymer densities $\phi_{A}$ and $\phi_{B}$. Next, we update the $\Omega_{+}$ field until the system satisfies the local incompressibility condition at every grid point. We repeat these two steps until the system reaches the equilibrium where the effective Hamiltonian is stabilized. We run $10^6$ iterations (each iteration contains to a single update of $\Omega_{-}$ and full update of corresponding $\Omega_{+}$) per simulation point until the system reaches equilibrium in this work.

\subsection{Description of anisotropic nanoparticles}

The cavity functions $\rho_{ext}$ for rod and disk NPs are modified here to account for the anisotropy of NPs in comparison to the isotropic cavity function for spheric NP previously described \cite{art:ykim}. For the rod NP, $\rho_{ext-rod}$ is a Gaussian function defined along the axis of the rod of length $L-D$ and standard deviation $0.5D$.  
$$ \rho_{ext-rod}(\textbf{r}) = \rho_{0} \exp( -  \frac{|\textbf{r} - \textbf{r}_{0}|^{2}}{2(0.5D)^{2}} )  $$
where $\textbf{r}_{0}$ is the closest point along the finite rod axis to $\textbf{r}$. In this way, we generate a soft rod NP of length $L$ and diameter $D$. For the disk NP, $\rho_{ext-disk}$ is a Gaussian function defined in plane of a disc of diameter $D-T$ with standard deviation $0.5T$.
$$ \rho_{ext-dis}(\textbf{r}) = \rho_{0} \exp( -  \frac{|\textbf{r} - \textbf{r}_{0}|^{2}}{2(0.5T)^{2}} )  $$
where $\textbf{r}_{0}$ is the closest point in the disk plane to $\textbf{r}$. In this way, we can generate a soft disk NP with diameter $D$ and thickness $T$.

To simulate such complex NP, an extremely refined grid is required to describe the NP correctly. Since the computational cost for a SCFT simulations directly scales with the number of chosen grid of simulation box, grid refinement of the entire simulation box is not an ideal solution. Instead, local grid refinement is required where the complex NPs exist to obtain better calculation accuracy and faster simulation execution. Such refinement is not possible for the traditional pseudo-spectral (PS) method for the diffusion-like equation solver because the simulation box has to be periodic. To solve this problem, we recently adapted the Lattice Boltzmann Method (LBM) to solve the diffusion-like equation used in SCFT simulations \cite{art:hchen}. In this study, the refined grid region is defined as a three dimensional box having as $2.0R_{g} \times 2.0R_{g} \times 2.0R_{g}$ to contain all field information of $\rho_{ext}$ which represents the rod and disk NP. Refinement of the fine grid is four times higher than coarse grid to reduce energy fluctuation of the molecule resulting from 3d rotation to the order of $10^{-6}$.

We set $\chi N$ to 20 with $f = 0.7$ where the minority block (B-block) forms hexagonally closed-packed cylinders in a bulk system. Each desired defect is designed by using the same concept as the templated-assembly of BCPs. By fixing the $\Omega_{-}$ field at desired locations, we can model the preferential wetting of one block of the BCP at certain grid points \cite{art:mickiewicz}. Then we optimized the system dimensions $L_{x}, L_{y}, L_{z}$ to minimize the free energy of the system with a desired defect shape. Using this method, we could generate X, T, and Y shape defects. For each defect, we plot the free energy of the system as a function of 3d rotation angle of director vector of rod and disk NP. We designed the NP to be highly attractive to the B-block but repulsive to the A-block by setting $\gamma_{A}$ to -40 and $\gamma_{B}$ to +40. 

\section{Results and Discussion}

\subsection{Case of normal cylinder without defect}

Before we study the mixture of rod and disk NPs with BCPs containing defects, we first simulated the system having a rod or disk NP within the minority block region of the defect-free cylinder phase of BCPs. First, we simulated the rod molecule with length $L$ and diameter $D$ (See Figure 1-a). The red arrow represents the director vector of the rod molecule which is rotated by an angle $\theta$ from the axis perpendicular to the cylinder axis (See Figure 1-b). In Figure 2, we plot the free energy of the system as a function of $\theta$, $L$, and $D$.

\begin{figure}[htbp]
\begin{center}
\includegraphics[scale=0.6]{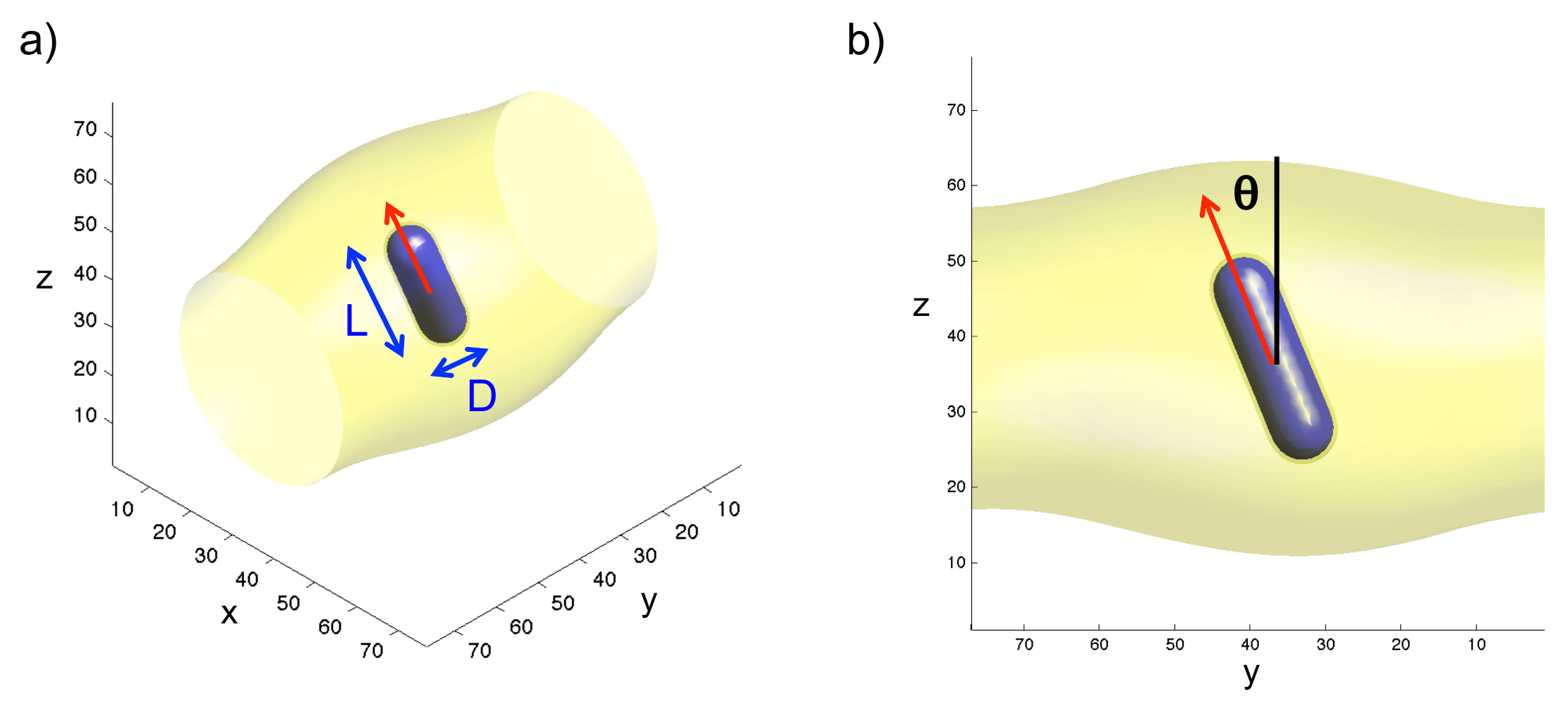}
\caption{a) Simulation picture of a self-assembled cylinder phase of BCPs containing a rod NP within the minority phase. Transparent yellow contour indicates region of constant volume density of BCPs with $\phi_{B}$ =  $\phi_{A}$ = 0.5. The red arrow represents the director vector of the rod NP. $L$ and $D$ represent the length and diameter of the rod respectively. b) We rotate the rod with angle $\theta$ with respect to the axis perpendicular to the cylinder axis to plot the free energy.  }
\end{center}
\end{figure}

\begin{figure}[htbp]
\begin{center}
\includegraphics[scale=0.6]{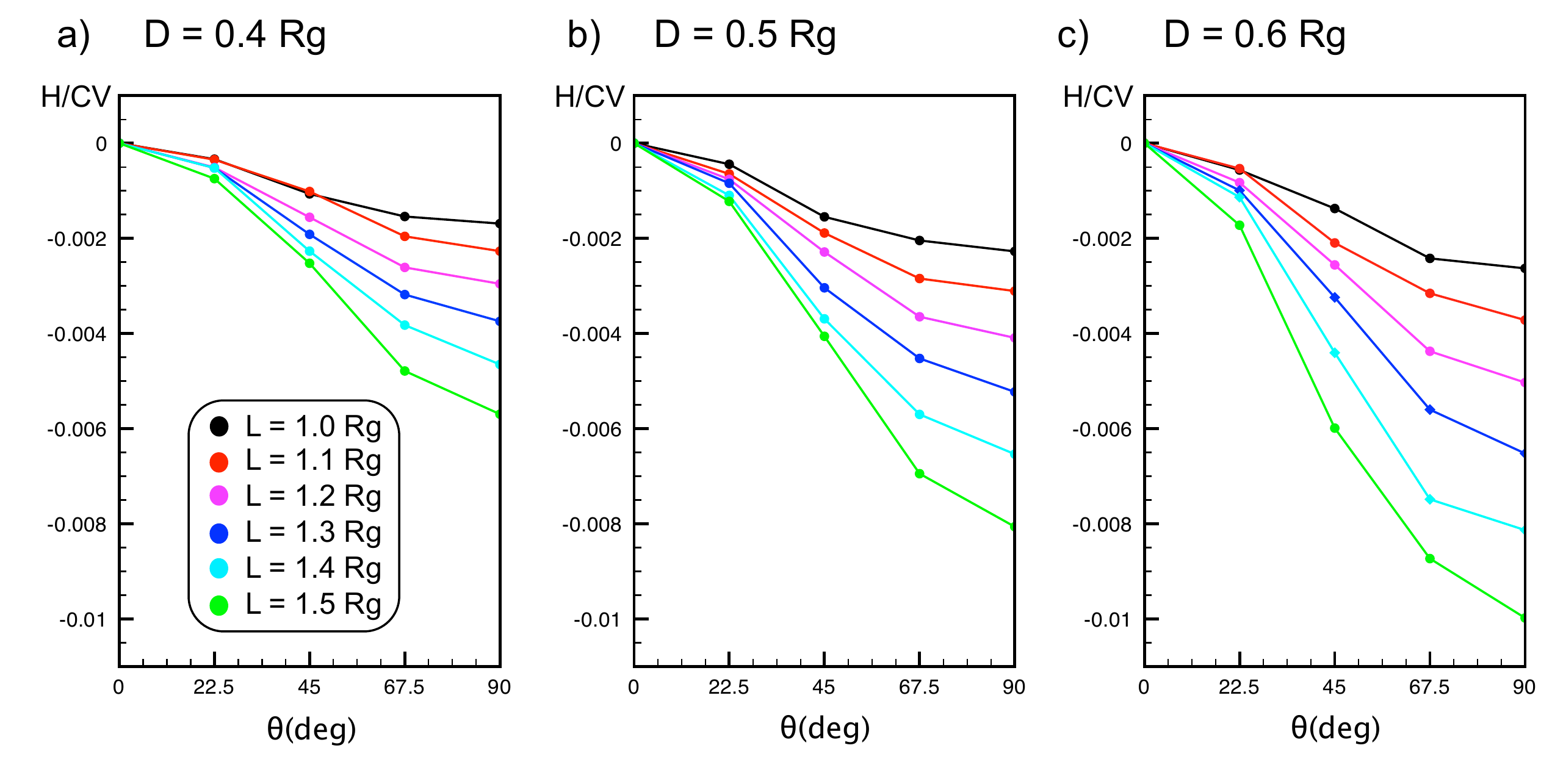}
\caption{Plot of the reduced free energy H/CV as a function of the rotation angle $\theta$ with different lengths $L$ and diameter $D$ of the rod NP : a) $D = 0.4 R_{g}$ b) $D = 0.5 R_{g}$ c) $D = 0.6R_{g}$. Different colors represent different lengths of the rod : 
$L = 1.0R_{g}$ (black), $L = 1.1R_{g}$ (red), $L = 1.2R_{g}$ (purple), $L = 1.3R_{g}$ (blue), $L = 1.4R_{g}$ (light blue), and $L = 1.5R_{g}$ (light green). }
\end{center}
\end{figure}

From Figure 2, the rod NP has the highest free energy when the director vector is perpendicular to the cylinder axis ($\theta = 0^ {\circ}$). If the rod is aligned perpendicular to the cylinder axis, the interface between the majority and minority blocks should be curved toward the director vector. This results in an increase of both the stretching energy and interfacial energy. However, if the rod molecule is aligned parallel to the cylinder axis ($\theta = 90^ {\circ}$), the occupation of the rod component in the vicinity of the polymer ends is maximized. This results in the minimization of both the stretching energy and the interfacial energy of the block copolymer, and the free energy shows a corresponding minimum for every value of $L$ and $D$ of the rod NP at $\theta = 90^ {\circ}$. Also, this minimization of the free energy is amplified if we increase the rod length $L$ and diameter $D$ because more rod volume will occupy the cylinder axis. We can clearly see this feature in Figure 2. When we increase $L$ and $D$, the free energy has a lower minimum at $\theta = 90^ {\circ}$. 

A disk NP having $D$ as diameter and $T$ as thickness was also simulated within the defect-free cylinder matrix (See Figure 3a). The red arrow represents the director vector of the disk which is rotated by an angle $\theta$ with respect to the axis perpendicular to the cylinder axis (See Figure 3-b). In Figure 4, we plot the free energy of the system as a function of $\theta$, $D$, and $T$. 

\begin{figure}[htbp]
\begin{center}
\includegraphics[scale=0.6]{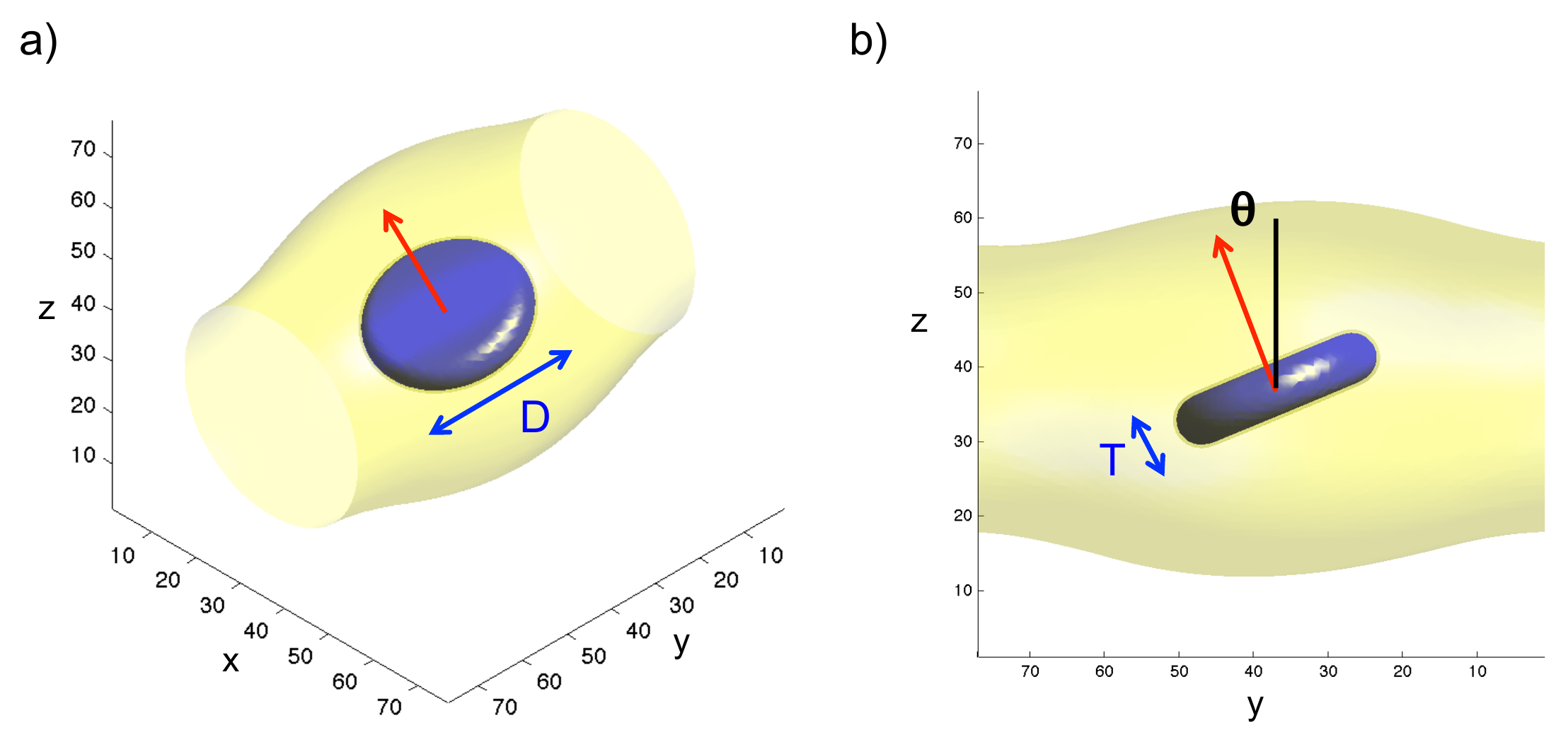}
\caption{a) Simulation picture of a self-assembled cylinder phase of BCPs containing a disk NP within the minority phase. Transparent yellow contour indicates region of constant volume density of BCPs with $\phi_{B}$ =  $\phi_{A}$ = 0.5. The red arrow represents the director vector of the disk NP and $D$ represents its diameter. The disk thickness $T$ is described in Figure 3b. b) We rotate the disk with angle $\theta$ with respect to the axis perpendicular to the axis of the cylinder to plot the free energy.  }
\end{center}
\end{figure}

\begin{figure}[htbp]
\begin{center}
\includegraphics[scale=0.6]{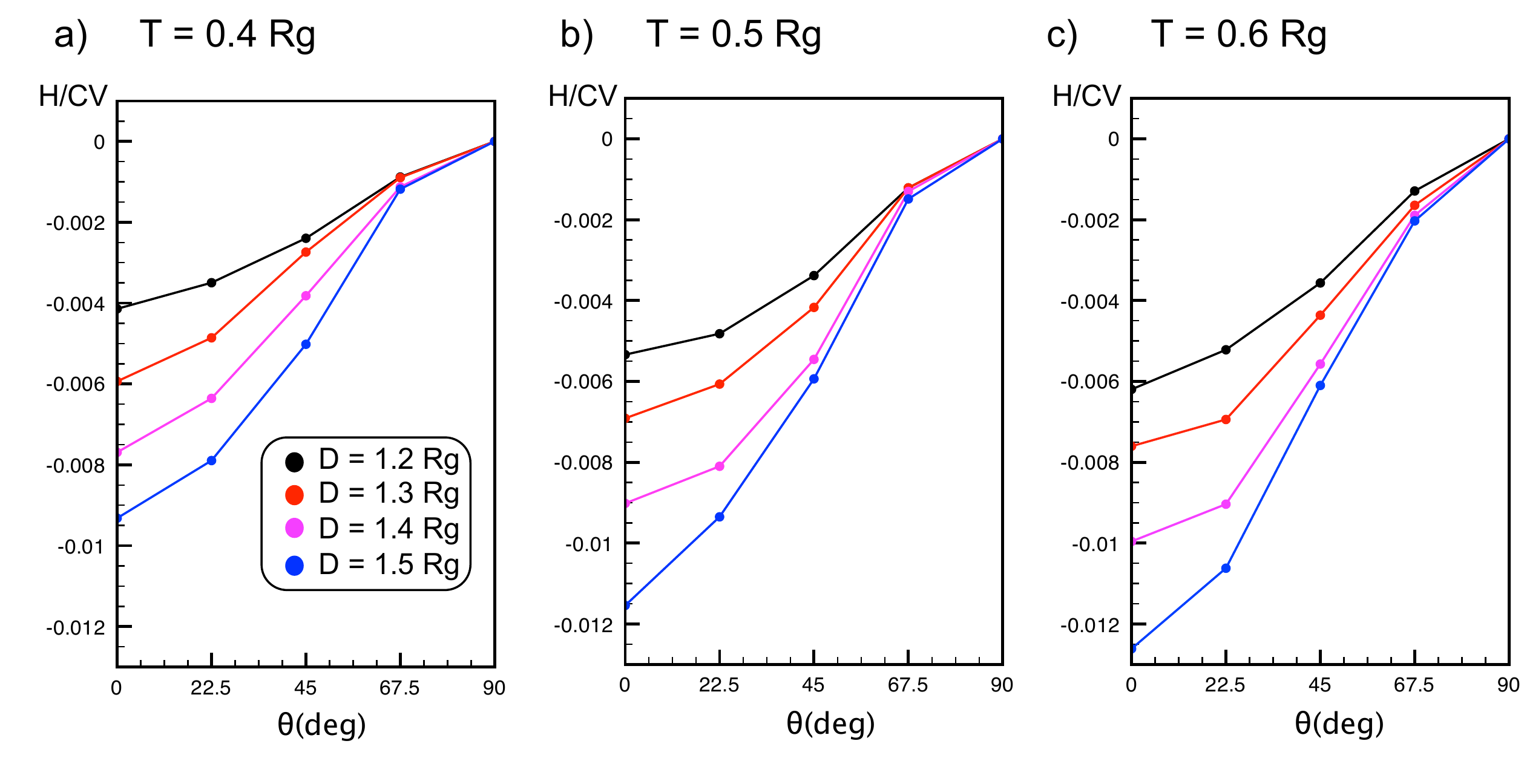}
\caption{Plot of the reduced free energy H/CV as a function of the rotation angle $\theta$ with different diameters $D$ and thicknesses $T$ of the disk NP : a) $T = 0.4 R_{g}$ b) $T = 0.5 R_{g}$ c) $T = 0.6R_{g}$. Different colors represent different thicknesses of the disk : 
$L = 1.2R_{g}$ (black), $L = 1.3R_{g}$ (red), $L = 1.4R_{g}$ (purple), and $L = 1.5R_{g}$ (blue). }
\end{center}
\end{figure}

In contrast to the rod case, the free energy minimum is achieved when the director vector of the disk is perpendicular to the cylinder axis ($\theta =0^ {\circ}$). This configuration of the disk NP maximizes the occupation of the disk at the axis of the cylinder where the polymer ends exist, minimizing the stretching energy of the block copolymer. Intuitively, we also can derive that the larger and thicker disk decreases the free energy more by reducing $\theta$ to $0^ {\circ}$. We can clearly see this feature in Figure 4. When we increase $D$ and $T$, the free energy has a lower minimum at $\theta = 0^ {\circ}$. 

\subsection{Case of X-shape defect}

From the defect-free cylinder case, we established that a free energy minimum is realized when most of the volume of the molecule occupies the cylinder axis where the ends of BCPs exist. However, when a defect is introduced to the system, the free energy well is constructed at the defect center. The stiffness of the free energy is different when we move the external particle in different directions from the defect center and we found that it is due to the different curvature of interface of BCPs which isotropic (spheric) NP is faced at \cite{art:ykim}. Therefore, different from spheric NP case, we can expect that the anisotropic feature of the rod and disk NPs results in preferred orientations of their director vectors near the defect center. From this section, we study the free energy landscape as a function of 3D rotation of director vector of rod and disk NPs at the center of three model defects.

The first case considered is the rod and disk NPs within the X-shape defects. Since the size of NP has been relevant to the magnitude but not the tendency of free energy landscape from defect-free cylinder case, we only consider the single size of rod and disk NP ( $D = 0.4R_{g}$, $L = 1.2R_{g}$ for rod and $D = 1.2R_{g}$, $T = 0.4R_{g}$ for disk). In Figure 5,  we show the rod and disk NPs are located at the center of X-shape defect structure of BCPs. Red arrow in Figure 5 represents the director vector of each NP and we plot the free energy as a function of director vector at the defect center in Figure 6 for rod and Figure 7 for disk respectively.

\begin{figure}[htbp]
\begin{center}
\includegraphics[scale=0.5]{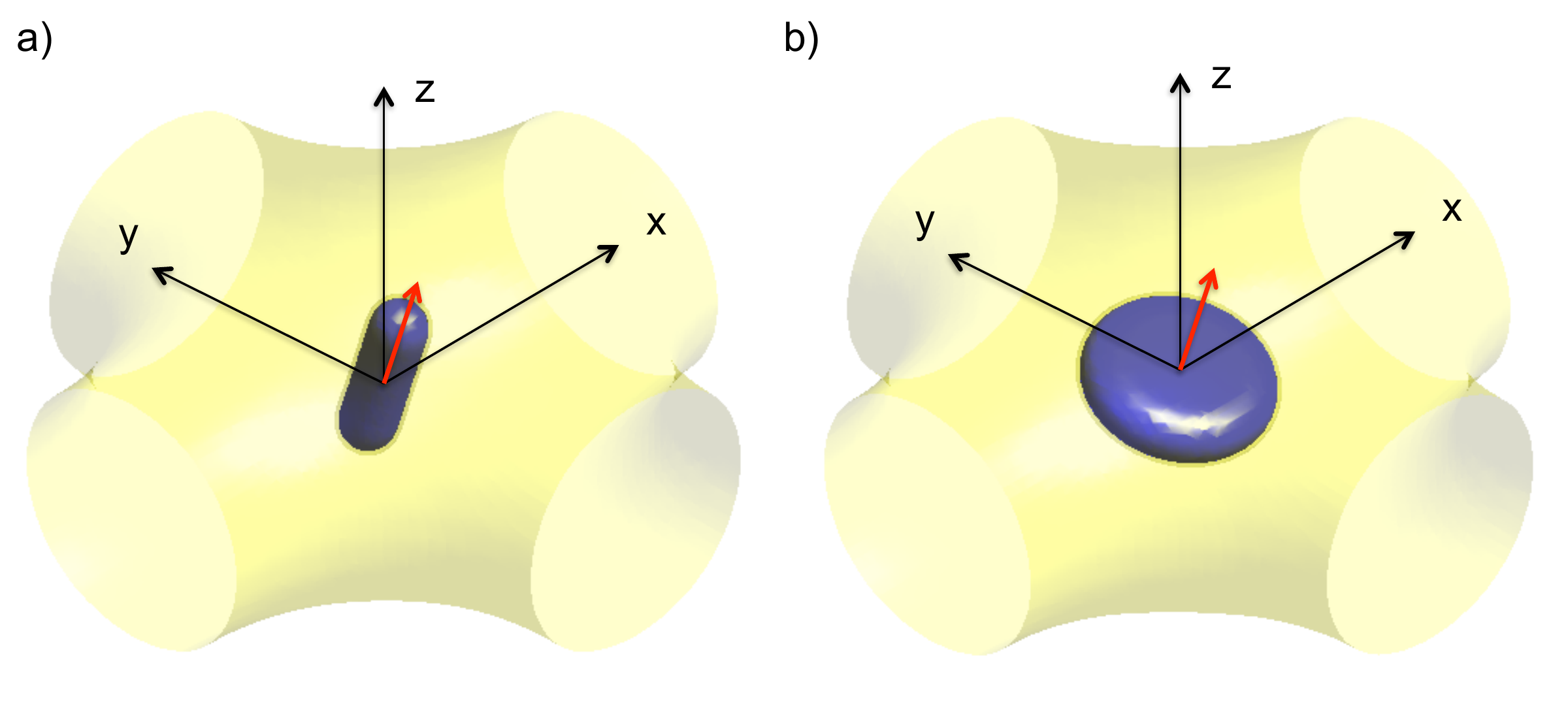}
\caption{a) Simulation picture of a self-assembled structure of a X-shape defect of BCPs containing a) rod and b) disk NP confined near the defect center in the minority phase. Transparent yellow contour indicates region of constant volume density of BCPs with $\phi_{B}$ =  $\phi_{A}$ = 0.5. The red arrow represents the director vector of each NP.}
\end{center}
\end{figure}

\begin{figure}[htbp]
\begin{center}
\includegraphics[scale=0.5]{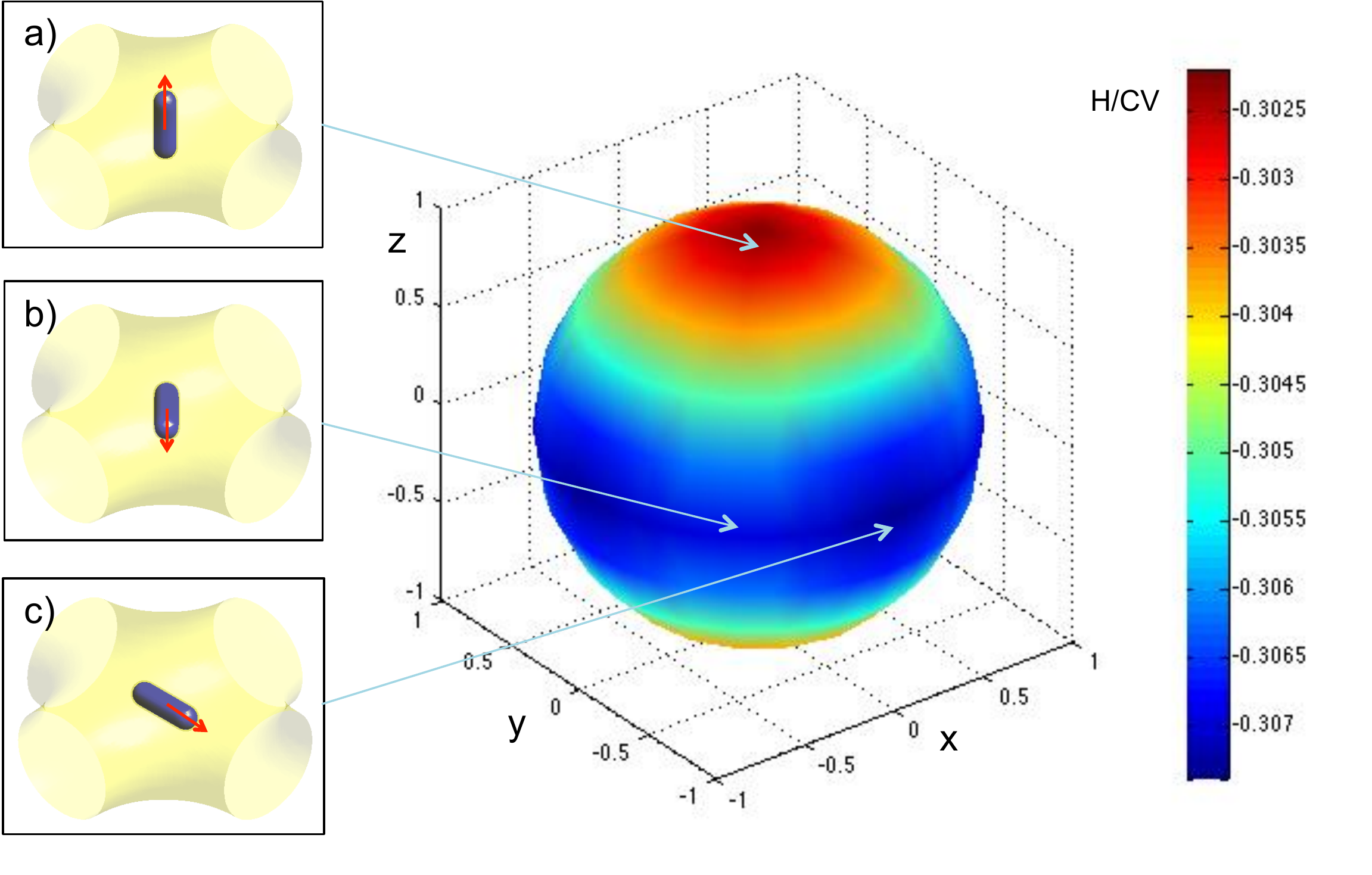}
\caption{a) Plot of free energy as a function of director vector of rod NP at the center of X-shape defect of BCPs. We colored reduced free energy $H/CV$ at the point where director vector designate from the origin. We also show some snapshots of important points in Figure 6-a), 6-b) and 6-c). }
\end{center}
\end{figure}

\begin{figure}[htbp]
\begin{center}
\includegraphics[scale=0.5]{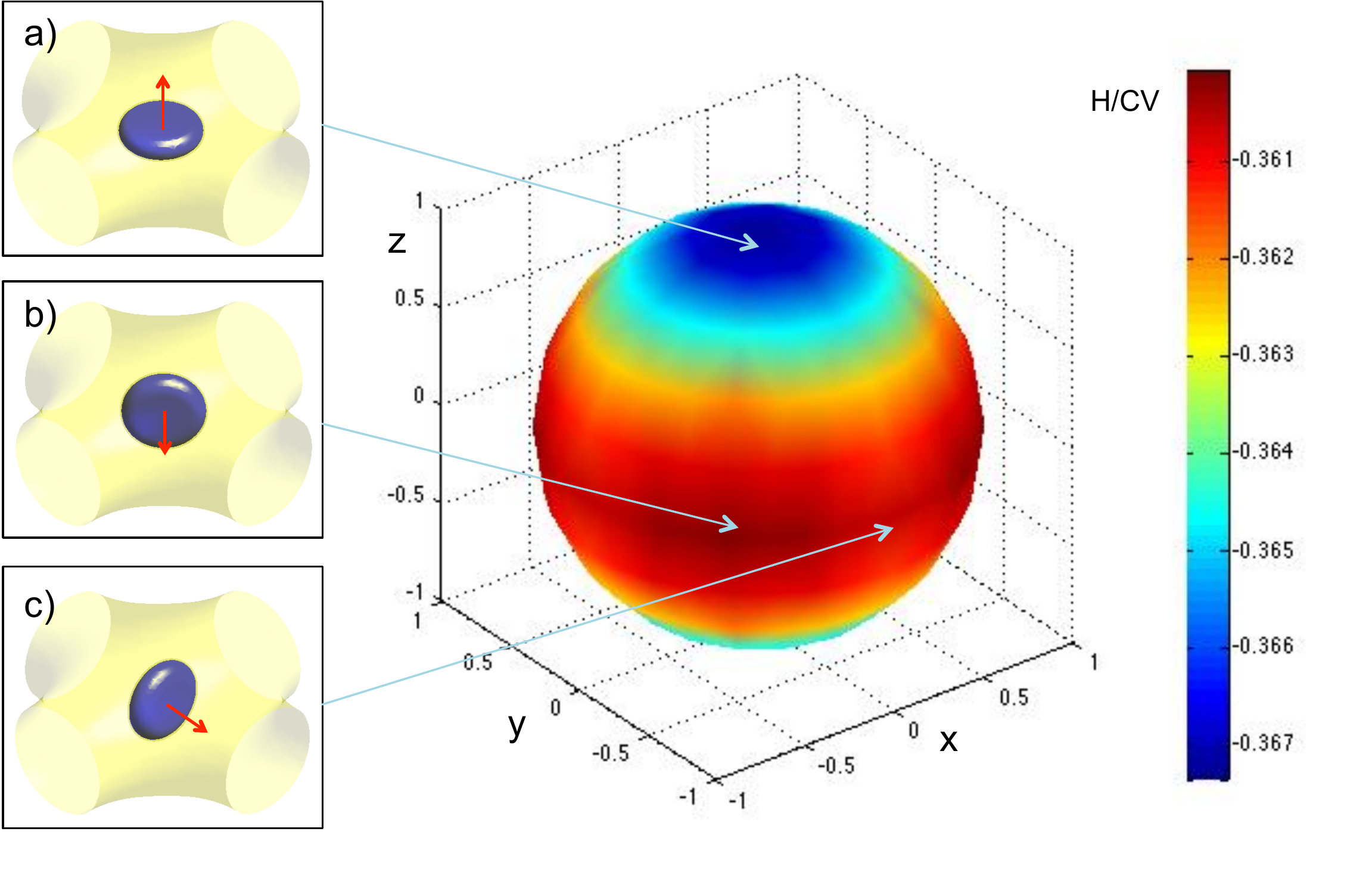}
\caption{a) Plot of free energy as a function of director vector of disk NP at the center of X-shape defect of BCPs. We colored reduced free energy $H/CV$ at the point where director vector designate from the origin. We also show some snapshots of important points in Figure 7-a), 7-b) and 7-c). }
\end{center}
\end{figure}

From Figure 6-a, the free energy maximum is achieved when the director vector of the rod is perpendicular to the two cylinder axes where the chain stretching and interfacial area is maximized. For rod NP, free energy gets lower when we rotate the director vector toward the plane which contains both cylinder axes of BCPs as seen in Figure 6-b and 6-c. However, aligning the rod to one of two cylinder axes (Figure 6-c) results lower free energy compared to the configuration where the director vector is aligned toward the interface of BCPs (Figure 6-b) due to the extra increment of interfacial area of BCPs. Therefore, we can conclude that the alignment of the rod to x or y axis is equilibrated configuration of the rod within the X-shape defect of BCPs. For the disk NP, different from the rod case, free energy is maximize when the director vector of the disk shares the same plane with two axes of BCP cylinders due to the large interfacial area stretching to z-direction(Figure 7-b and 7-c). When the director vector of the disk is perpendicular to the two axes of cylinders where the volume of the disk shares both cylinder axes (Figure 7-a), the minimization of free energy is achieved from the minimized stretching of polymer chains and interfacial area.

\subsection{Case of T-shape defect}

The T-shape defect of BCPs is also simulated with the rod and disk NPs. Similar to the X-shape defect case, we rotate the director vector of the rod and disk at the defect center of T-shape BCP cylinders as shown in Figure 8-a and 8-b. In Figure 9, we plot the free energy the rod within the T-shape defect as a function of director vector. We also plot the free energy of the disk NP within the T-shape defect of BCPs in Figure 10. 

\begin{figure}[htbp]
\begin{center}
\includegraphics[scale=0.5]{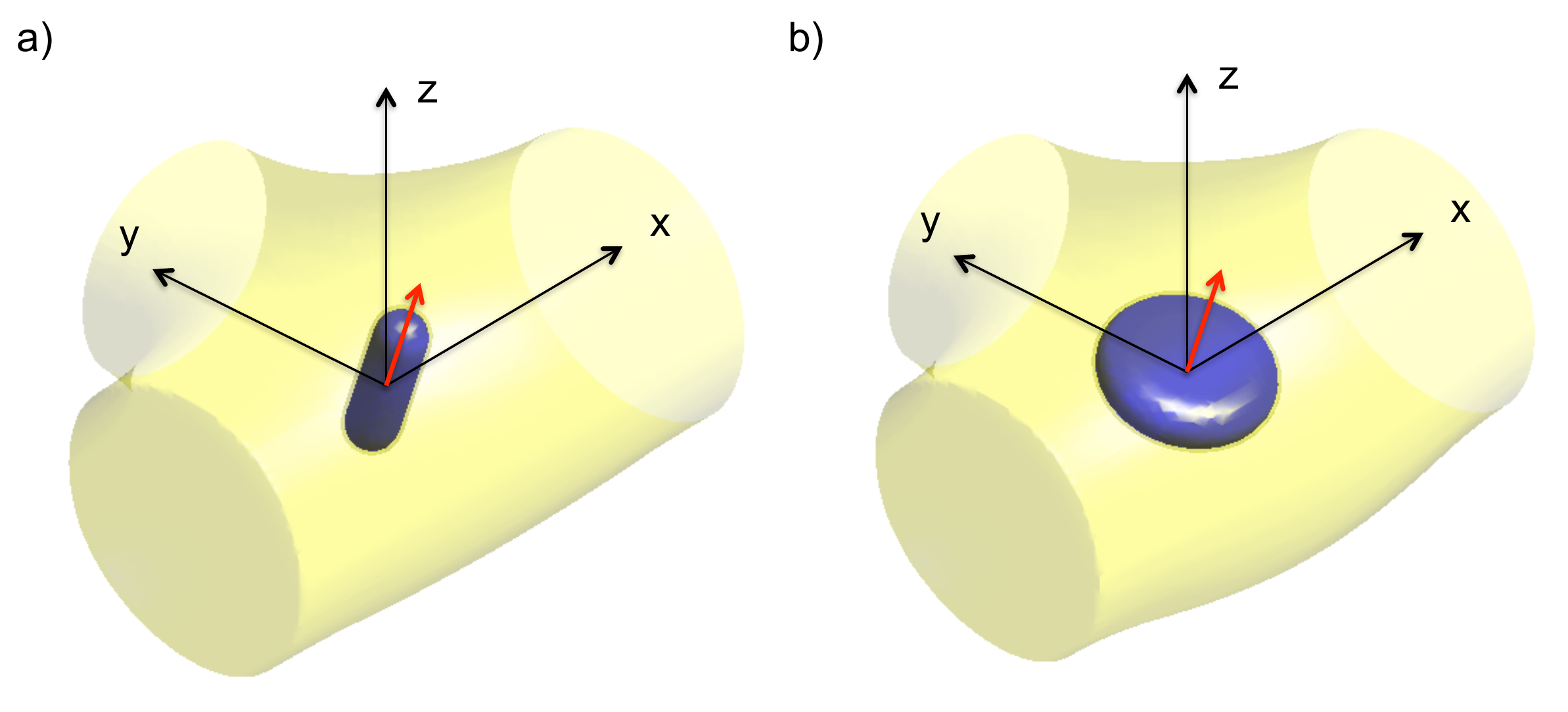}
\caption{a) Simulation picture of a self-assembled structure of a T-shape defect of BCPs containing a) rod and b) disk NP confined near the defect center in the minority phase. Transparent yellow contour indicates region of constant volume density of BCPs with $\phi_{B}$ =  $\phi_{A}$ = 0.5. The red arrow represents the director vector of each NP.}
\end{center}
\end{figure}

\begin{figure}[htbp]
\begin{center}
\includegraphics[scale=0.5]{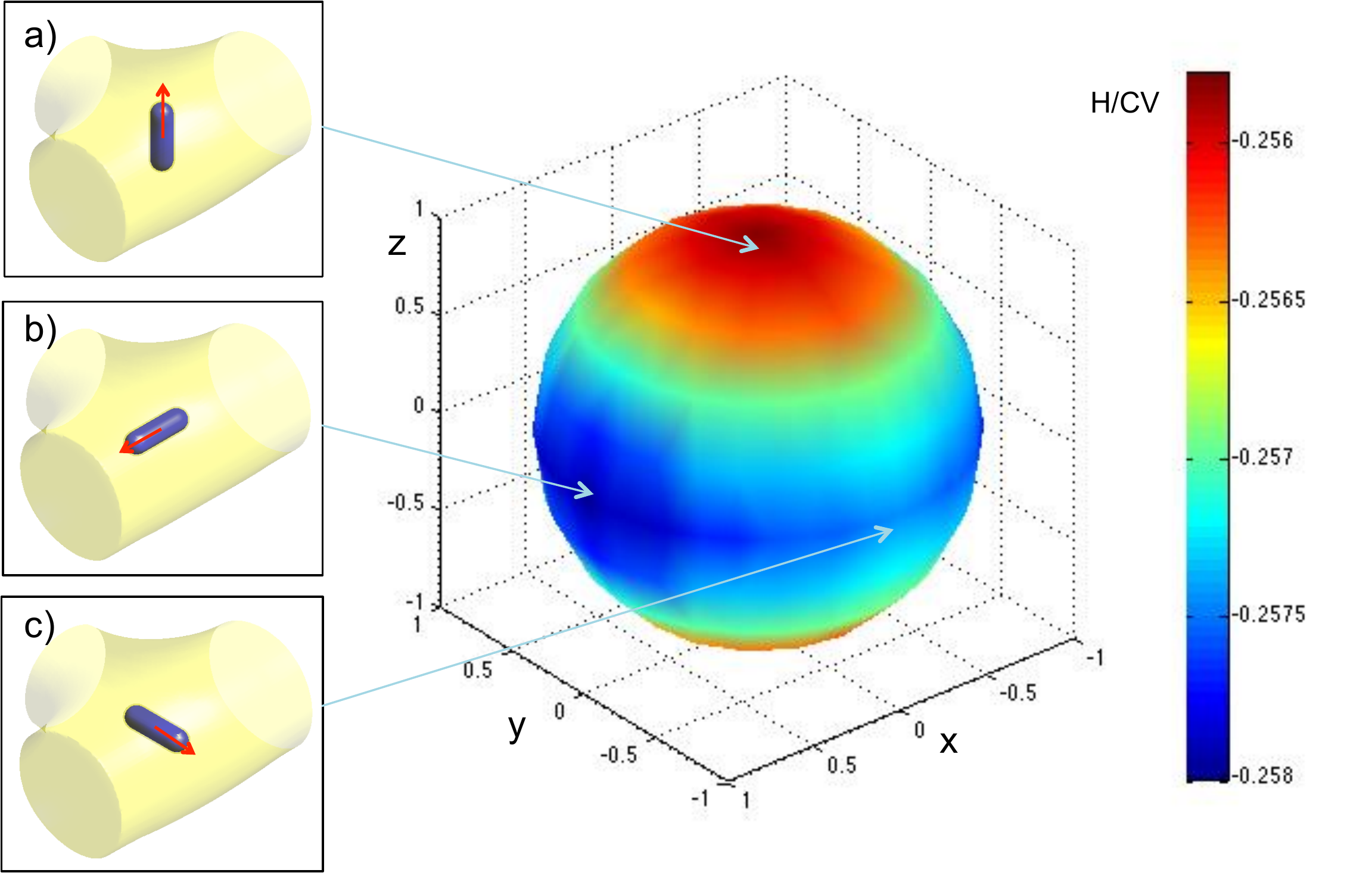}
\caption{a) Plot of free energy as a function of director vector of rod NP at the center of T-shape defect of BCPs. We colored reduced free energy $H/CV$ at the point where director vector designate from the origin. We also show some snapshots of important points in Figure 9-a), 9-b) and 9-c). }
\end{center}
\end{figure}

\begin{figure}[htbp]
\begin{center}
\includegraphics[scale=0.5]{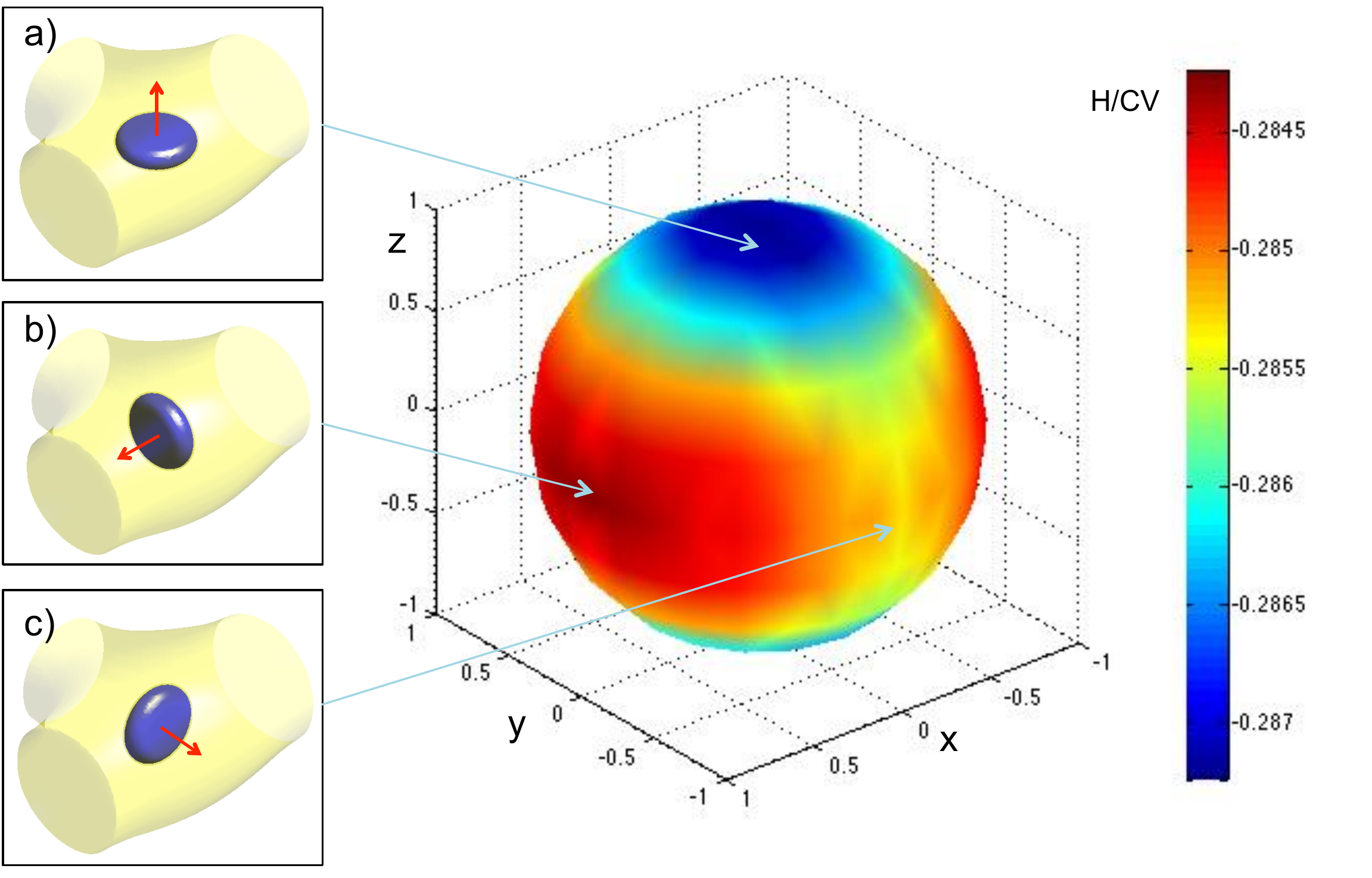}
\caption{a) Plot of free energy as a function of director vector of disk NP at the center of T-shape defect of BCPs. We colored reduced free energy $H/CV$ at the point where director vector designate from the origin. We also show some snapshots of important points in Figure 10-a), 10-b) and 10-c). }
\end{center}
\end{figure}

For the X-shape defect, the rod maximize the volume occupied along the cylinder axis of BCPs by aligning with either the x or y axis. However, for the T-shape defect, if the rod is aligned toward the axial direction of the cylinder aligned along the y axis as shown in Figure 9-c, the rod molecule will experience more curved interface of BCPs. Therefore, the director vector of the rod prefers to be aligned with the x axis at the defect center in a more energetically favorable configuration (Figure 9-b). The free energy maximum is achieved when the director vector of the rod is perpendicular to two cylinder axes of cylinders at the defect center as shown in Figure 9-a. On the other hand, when the director vector of the disk NP is perpendicular to both cylinder axes (Figure 10-a), the free energy minimum is achieved because the disk shares two axes of cylinders at this configuration. As director vector of the disk rotate toward to the axial direction of cylinders (Figure 10-b and 10-c), the free energy is increased due to the stretching of polymers and interfacial area. Among those configurations, free energy is maximized when the director vector of the disk aligned toward x axis as shown in FIgure 10-b since the disk has faced highly curved interface of BCPs. 

\subsection{Case of Y-shape defect}

Finally, we simulate the rod and disk NPs with Y-shape defects. As shown in Figure 11, we rotate each director vector of rod and disk at the center of Y-shape defect structure. We also plot the free energy landscape as a function of the director vector of the rod and disk in Figure 12 and Figure 13 respectively.

\begin{figure}[htbp]
\begin{center}
\includegraphics[scale=0.5]{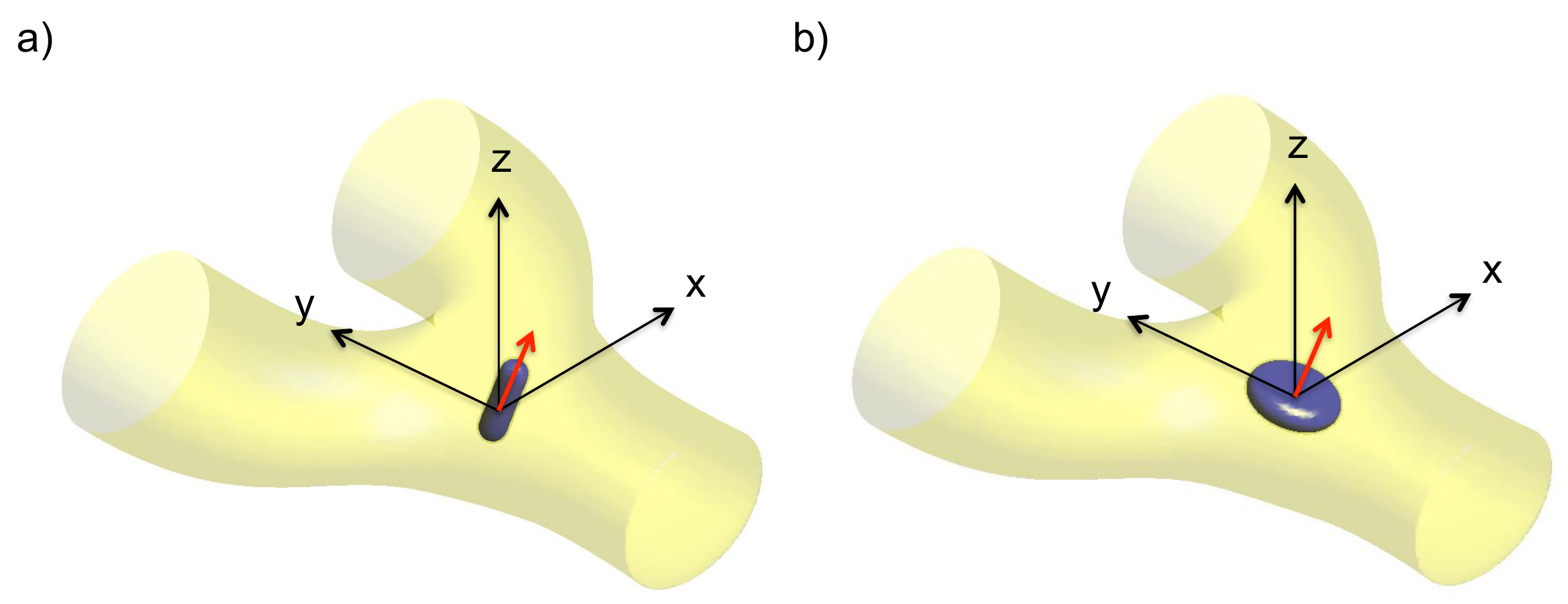}
\caption{a) Simulation picture of a self-assembled structure of a Y-shape defect of BCPs containing a) rod and b) disk NP confined near the defect center in the minority phase. Transparent yellow contour indicates region of constant volume density of BCPs with $\phi_{B}$ =  $\phi_{A}$ = 0.5. The red arrow represents the director vector of each NP.}
\end{center}
\end{figure}

\begin{figure}[htbp]
\begin{center}
\includegraphics[scale=0.5]{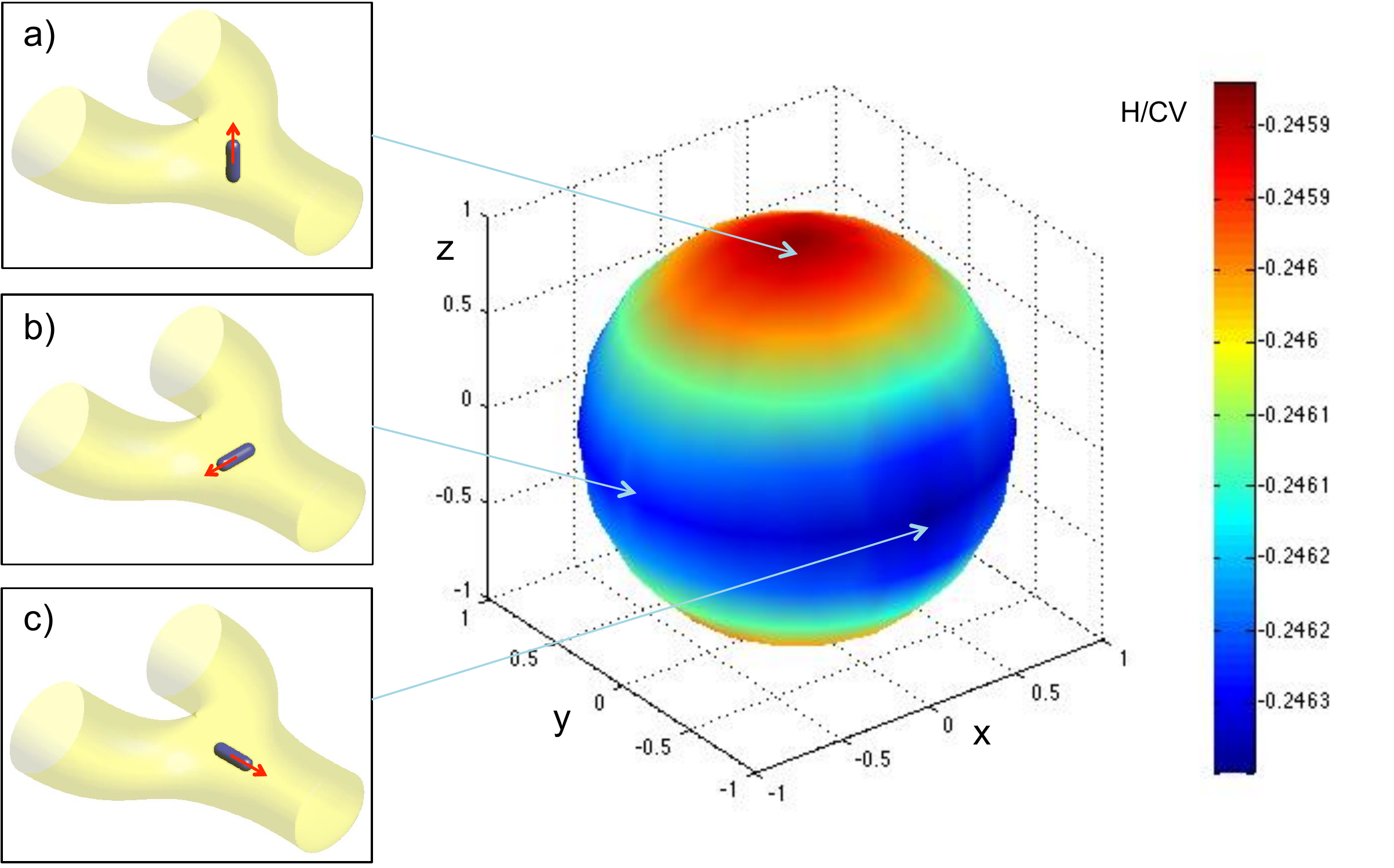}
\caption{a) Plot of free energy as a function of director vector of rod NP at the center of Y-shape defect of BCPs. We colored reduced free energy $H/CV$ at the point where director vector designate from the origin. We also show some snapshots of important points in Figure 12-a), 12-b) and 12-c). }
\end{center}
\end{figure}

\begin{figure}[htbp]
\begin{center}
\includegraphics[scale=0.5]{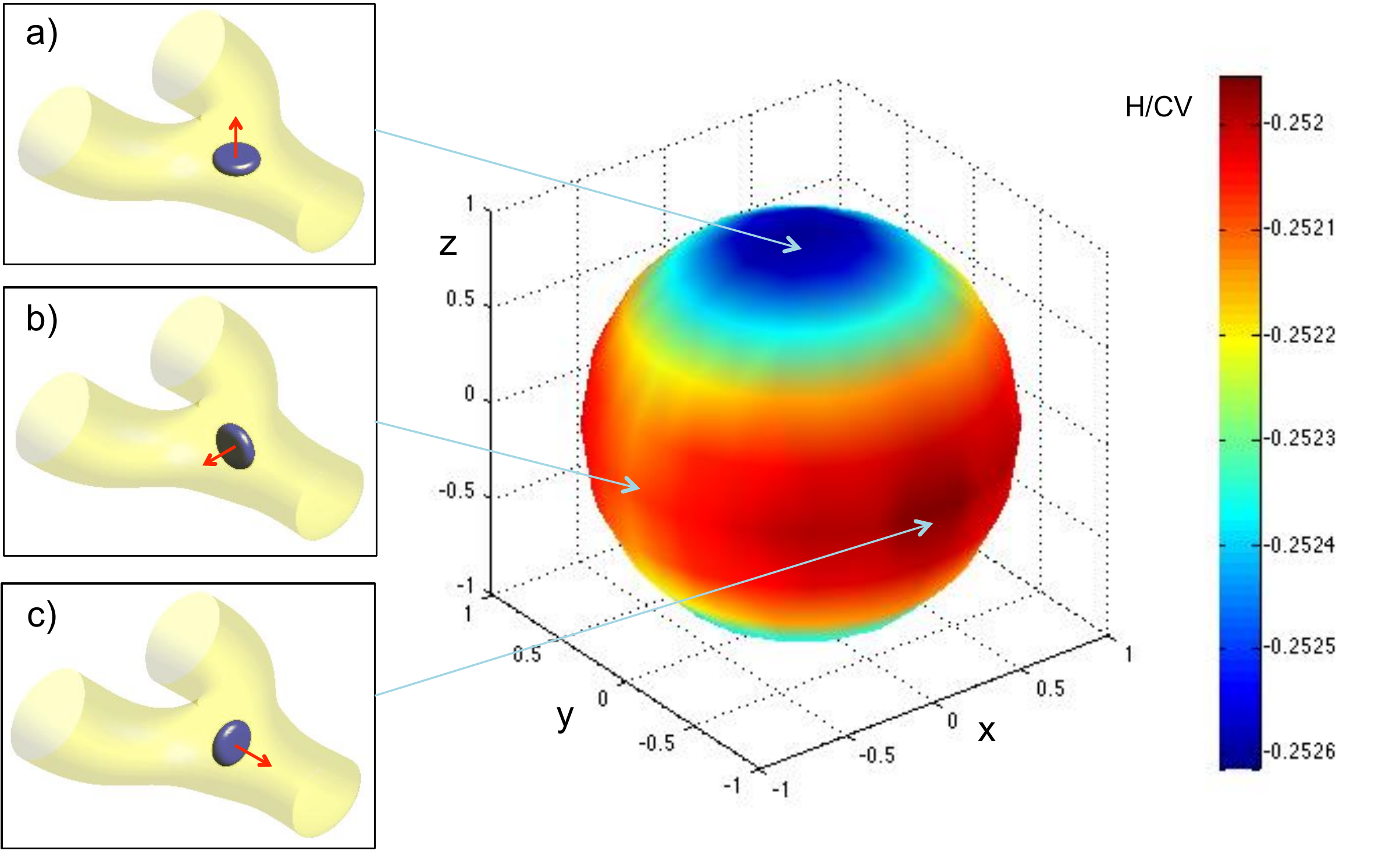}
\caption{a) Plot of free energy as a function of director vector of disk NP at the center of Y-shape defect of BCPs. We colored reduced free energy $H/CV$ at the point where director vector designate from the origin. We also show some snapshots of important points in Figure 13-a), 13-b) and 13-c). }
\end{center}
\end{figure}

Similar to the other defect structures of BCPs, the free energy maximum is achieved when the director vector of the rod is perpendicular to the three cylinder axes as shown in Figure 12-a. Since two other cylinders (aligned to the (+x,+y) direction and (-x,+y) direction) provide a highly curved interface of two blocks compare to the cylinder aligned to the (-y) direction, the extra stretching energy from highly curved interface prohibits the rod from aligning with those two axial directions. As a result, rod aligned to y-axis has the lowest free energy than any other configuration of this system as shown in Figure 12-b and Figure 12-c. As expected, for the disk NP, the free energy minimum is achieved when the director vector is aligned perpendicular to the axial directions of all cylinders (Figure 13-a). The free energy maxima is achieved when the director vector of the disk aligned toward the (-y) direction because the disk directly faces the highest curved interface of BCPs at this configuration as shown in Figure 13-c.

\section{Conclusions}

In this study, we investigated the behavior of rod and disk NP under the confinement of various defect structures within a BCP matrix. Self-Consistent Field Theory (SCFT) simulations with cavity function for NPs were used to describe the rod and disk NPs as an extra field excluding the surrounding polymer density. By solving the diffusion-like equation with the Lattice Boltzmann Method (LBM), we could use the local grid refinement to simulate the rod and disk NP with higher accuracy than possible with the pseudo-spectral method.

We first plotted the free energy of the rod and disk NP as a function of rotation angle within a defect-free cylinder phase of the BCPs. For the rod, the free energy minimum is achieved when the director vector is aligned with the cylinder axis. In this configuration, the rod molecule maximizes the volume occupancy of the cylinder axis where the polymer ends exist to minimize the stretching energy of the BCPs. This is also applied to the disk NP case, where the director vector of the disk should be perpendicular to the axial direction of cylinder for the maximum volume occupancy of the cylinder axis. For this configuration, the system reaches the free energy minimum. 

Based on the result from defect-free cylinder case, we also studied the rod and disk NPs confined within the X, T, and Y-shape defects of the BCPs. Unlike the defect-free cylinder case, here the curved interface of the BCPs add an extra stretching energy to the system. As a result, a higher energy barrier exists near the defect center when the NP approaches the highly curved interfaces. For the anisotropic NPs, this determines the specific configuration of the director vector of NP within the defect structure of the BCPs. For the X-shape defect case, the director vector of the rod is aligned with one of two cylinders at the defect center (cylinder junction) due to the symmetry of the X-shape defect. The director vector of the rod in both the T-shape and Y-shape defect is aligned with the x axis (Figure 9-b) and y axis (Figure 12-c) respectively to minimize the extra chain stretching from the curved interface. For the disk NP, the director vector is perpendicular to the defect plane which shares every axial direction of cylinders to minimize the free energy for all types of defects. In this configuration, the disk NP can occupy all cylinder axes equally to maximize the volume coverage by the disk itself. Since the chain stretching energy is important factor for the preferred orientation of NPs, we expect that BCPs contain semi-flexible polymer will enhance the confinement effect. There results give a better understanding of the effect of confinement on the preferred configuration of complex and anisotropic NPs blended with BCPs. Furthermore, we believe that we can suggest the design principles of complex systems by controlling the alignment of anisotropic NPs within the BCP scaffold for various applications.

\begin{acknowledgement}
This work was supported by the U.S. Department of Energy, Office of Basic Energy Sciences, Division of Materials Science and Engineering under Award No. \#ER46919.
\end{acknowledgement}


\begin{thebibliography}{9}

\bibitem{art:alivisatos}
A. P. Alivisatos, K. P. Johnson, X. Peng, T. E. Wilson, C. J. Loweth, M. P. Bruchez and P. G. Schultz, {\it Nature} 382, 609 (1996).

\bibitem{art:mirkin}
C. A. Mirkin, R. L. Letsinger, R. C. Mucic, and J. J. Storhoff, {\it Nature} 382, 607 (1996). 

\bibitem{art:yin}
Y. Yin and A. P. Alivisatos, {\it Nature} 437, 664 (2004).

\bibitem{art:nykypanchuk}
D. Nykypanchuk, M. M. Maye, D. van der Lelie, and O. Gang, {\it Nature} 451, 549 (2008).

\bibitem{art:devries}
G. A. DeVries, M. Brunnbauer, Y. Hu, A. M. Jackson, B. Long, B. T. Neltner, O. Uzun, B. H. Wunsch, and F. Stellacci, {\it Science} 315, 358 (2007).

\bibitem{art:choi}
C. L. Choi and A. P. Alivisatos, {\it Annu. Rev. Phys. Chem} 61, 369 (2010).

\bibitem{art:xu}
H. Xu, J. Aizpurua, M. Kall, and P. Apell, {\it Phys. Rev. E} 62, 4318 (2000).

\bibitem{art:chen}
C.-F. Chen, S.-D. Tzeng, H.-Y. Chen, K.-J. Lin, and S. Gwo, {J. Am. Chen. Soc.} 130, 824 (2008).

\bibitem{art:mulvihill}
M. J. Mulvihill, X. Y. Ling, J. Henzie, and P. Yang, {\it J. Am. Chem. Soc.} 132, 268 (2009).

\bibitem{art:mukherjee}
B. Mukherjee and M. Mukherjee, {\it Appl. Phys. Lett.} 94, 173510 (2009).

\bibitem{art:park}
Y-S Park, S. Y. Lee, and J. S. Lee, {\it IEEE Electron Device Lett.} 31, 1134 (2010).

\bibitem{art:cui}
P. Cui, S. Seo, J. Lee, L. Wang, E. Lee, M. Min, and H. Lee, {\it ACS Nano} 5, 6826 (2011).

\bibitem{art:pattantyus-abraham}
A. G. Pattantyus-Abraham, I. J. Kramer, A. R. Barkhouse, X. Wang, G. Konstantatos, R. Debnath, L. Levina, I. Raabe, M. K. Nazeeruddin, M. Gratzel, and E. H. Sargent, {\it ACS Nano} 4, 3374 (2010).

\bibitem{art:ip}
A. H. Ip, S. M. Thon, S. Hoogland, O. Voznyy, D. Zhitomirsky, R. Debnath, L. Levina, L. R. Rollny, G. H. Carey, A. Fischer, K. W. Kemp, I. J. Kramer, Z. Ning, A. J. Labelle, K. W. Chou, A. Amassian, and E. H. Sargent, {\it Nature Nanotechnology} 7, 577 (2012).

\bibitem{art:jean}
J. Jean, S. Chang, P. R. Brown, J. J. Cheng, P. H. Rekemeyer, M. G. Bawendi, S. Gradecak, V. Bulovic, {\it Adv. Mater} 25, 2790 (2013).

\bibitem{art:nie}
Z. Nie, A. Petukhova, and E. Kumacheva, {\it Nature Nanotechnology}, 5, 15 (2010).

\bibitem{art:ng}
K. C. Ng, I. B. Udagedara, I. D. Rukhlenko, Y. Chen, T. Yang, M. Premaratne, and W. Cheng, {\it ACS Nano}, 6, 925 (2012)

\bibitem{art:ritter}
K. A. Ritter and J. W. Lyding, {\it Nature Mater.}, 8, 235 (2009).

\bibitem{art:zhuo}
S. Zhuo, M. Shao, and S-T Lee, {\it ACS Nano}, 6, 1059 (2012).

\bibitem{art:jlee}
J. Lee, K. Kim, W. I. Park, B-H Kim, J. H. Park, T-H Kim, S. Bong, C-H Kim, G. Chae, M. Jun, Y. Hwang, Y. S. Jung, and S. Jeon, {\it Nano Lett.}, 12, 6078 (2012).

\bibitem{art:stengl}
V. Stengl and J. Henych, {\it Nanoscale}, 5, 3387 (2013).

\bibitem{art:gopalakrishnan}
D. Gopalakrishnan, D. Damien, and M. M. Shaijumon, {\it ACS Nano}, Accepted (2014). 

\bibitem{art:leibler}
L. Leibler, {\it Macromolecules} 13, 1602 (1980).

\bibitem{art:bates}
F. S. Bates and G. H. Fredrickson, {\it Annu. Rev. Phys. Chem.} 41, 525 (1990).

\bibitem{art:cpark}
C. Park, J. Yoon, and E. L. Thomas, {\it Polymer} 44, 6760 (2003).

\bibitem{art:chiu}
J. J. Chiu, B. J. Kim, E. J. Kramer, and D. J. Pine, {\it J. Am. Cxhem. Soc.} 127, 3036 (2005).

\bibitem{art:bkim}
B. J. Kim, J. Bang, C. J. Hawker, and E. J. Kramer, {\it Macromolecules} 39, 4108 (2006).

\bibitem{art:thompson}
R. B. Thompson, V. V. Ginzburg, M. W. Matsen, and A. C. Balazs, {\it Science} 292, 2469 (2001). 

\bibitem{art:bockstaller1}
M. R. Bockstaller, Y. Lapetnikov, S. Margel, and E. L. Thomas, {\it J. Am. Chem. Soc.}  125, 5276 (2003).

\bibitem{art:kim}
S. H. Kim and E. W. Cochran, {\it Polymer} 52, 2328 (2011).

\bibitem{art:zhao}
Y. Zhao, K. Thorkelsson, A. J. Mastroianni, T. Schilling, J. M. Luther, B. J. Rancatore, K. Matsunaga, H. Jinnai, Y. Wu, D. Poulsen, J. M. J. Frechet, A. P. Alivisatos, and T. Xu, {\it Nature Mater.} 8, 979 (2009).

\bibitem{art:kao}
J. Kao, P. Bai, V. P. Chuang, Z. Jiang, P. Ercius, and T. Xu, {\it Nano Lett.} 12, 2610 (2012).

\bibitem{art:kao2}
J. Kao, S. J. Bai, Z. Jiang, D. H. Lee, K. Aissou, C. A. Ross, T. P. Russell, and T. Xu, {\it Adv. Mater.} 26, 2777 (2014).

\bibitem{art:thorkelsson}
K. Thorkelsson, J. H. Nelson, A. P. Alivisatos, and T. Xu, {\it Nano Lett.} 13, 4908 (2013).

\bibitem{art:listak}
J. Listak and M. R. Bockstaller, {\it Macromolecules} 39, 5820 (2006).

\bibitem{art:kang}
H. Kang, F. A. Detcheverry, A. N. Mangham, M. P. Stoykovich, K. C. Daoulas, R. J. Hamers, M. Muller, J. J. de Pablo, and P. F. Nealey, {\it Phys. Rev. Lett.} 100, 148303 (2008).

\bibitem{art:jkim}
J. Kim and P. F. Green, {\it Macromolecules} 43, 10452 (2010).

\bibitem{art:bita}
I. Bita, J. K. W. Yang, Y. S. Jeong, C. A. Ross, E. L. Thomas, and K. K. Berggren, {\it Science} 321, 939 (2008).

\bibitem{art:yang}
J. K. W. Yang, Y. S. Jeong, J. B. Chang, R. A. Mickiewicz, A. Alexander-Katz, C. A. Ross, and K. K. Berggren, {\it Nature Nanotechnology} 5, 256 (2010).

\bibitem{art:tavakkoli1}
A. Tavakkoli K. G, K. W. Gotrik, A. F. Hannon, A. Alexander-Katz, C. A. Ross, and K. K. Berggren {\it Science} 336, 1294 (2012).

\bibitem{art:tavakkoli2}
A. Tavakkoli K. G, A. F. Hannon, J. G. Son, B. Keller, A. Alexander-Katz, and C. A. Ross {\it Adv. Mater.} 24, 4249 (2012).

\bibitem{art:mickiewicz}
R. A. Mickiewicz, J. K. W. Yang, A. F. Hannon, Y. S. Jeong, A. Alexander-Katz, K. K. Berggren, and C. A. Ross, {\it Macromolecules} 43, 8290 (2010).

\bibitem{art:ykim}
Y. Kim, H. Chen, and A. Alexander-Katz, {\it Soft Matter} 10, 3284 (2014).

\bibitem{art:sides}
S. W. Sides, B. J. Kim, E. J. Kramer, and G. H. Fredrickson, {\it Phys. Rev. Lett.} 96, 250601 (2006). 

\bibitem{art:hchen}
H. Chen, Y. Kim, and A. Alexander-Katz, {\it J. Chem. Phys.} 138, 104123 (2013).





\end{thebibliography}
\end{document}